\documentclass[useAMS,usenatbib,usegraphicx]{mn2e}
\usepackage{amsmath,amssymb}
\usepackage{graphicx,mathptm} 
\usepackage{times} 
\usepackage{natbib}
\usepackage{bm}

\newcommand{\be}{\begin{equation}}
\newcommand{\ee}{\end{equation}}
\newcommand{\bea}{\begin{eqnarray}}
\newcommand{\eea}{\end{eqnarray}}

\title{Nonresonant Grain Acceleration in MHD Turbulence}

\author[H. Yan]{Huirong Yan\thanks{E-mail:
    yan@lpl.arizona.edu}
\\
University of Arizona, LPL, Steward Observatory, \& Dept of Physics, 1629 E University Blvd, Tucson, AZ 85721, USA
}
\begin{document}

\date{}

\pagerange{\pageref{firstpage}--\pageref{lastpage}} \pubyear{2009}

\maketitle

\label{firstpage}

\begin{abstract}
We discuss a new type of dust acceleration mechanism that acts in a turbulent magnetized medium. The magnetohydrodynamic (MHD) turbulence can accelerate grains through resonant as well as nonresonant interactions. We show that the magnetic compression provides higher velocities for super-Alfv\'enic turbulence and can accelerate an extended range of grains in warm media compared to gyroresonance. While fast modes dominate the acceleration for the large grains, slow modes can be important for sub-micron grains. We provide comprehensive discussion of all the possible grain acceleration mechanisms in interstellar medium. We show that supersonic velocities are attainable for Galactic dust grains. We discuss the consequence of the acceleration. The implications for extinction curve, grain alignment, chemical abundance, etc, are provided.
\end{abstract}


\section{INTRODUCTION}

Dust is an important constituent that is essential for heating and
cooling of the interstellar medium (ISM). It interferes with observations
in the optical range, but provides an insight to star-formation activity
through far-infrared radiation. It also enables molecular hydrogen
formation and traces the magnetic field via emission and extinction
polarimetry \citep[see][]{Hildebrand00, Lazarian07rev}. The basic
properties of dust (optical, alignment etc.) strongly depend on its
size distribution. The latter evolves as the result of grain collisions,
whose frequency and consequences (coagulation, cratering, shattering,
and vaporization) depend on grain relative velocities (see discussions
in \citealt*{Draine85, YLD04}, hereafter YLD04; \citealt{HY09}).

All these problems require our understanding of grain
motions in the interstellar medium. The interstellar medium is magnetized
and turbulent \citep[see][]{Arons75}. Although turbulence has been
invoked by a number of authors \citep[see][]{Kusaka70, Volk80, Draine85, Ossenkopf93, WeidenRuz94} to provide substantial grain relative motions, the turbulence 
they discussed was not magnetized. Dust grains are charged, and their interactions with MHD turbulence is different from the hydrodynamic case.

In \citet{LY02} and Yan \& Lazarian (2003, henceforth YL03), we considered acceleration of charged dust by the MHD turbulence. In particular, \citet{LY02} applied the theory of Alfv\'{e}nic
turbulence (\citealt{GS95}, henceforth GS95, see \citealt*[][for a review]{CLV_lecnotes} ) to grain acceleration through gaseous
drag;
YL03 identified a new mechanism for grain acceleration through gyroresonance. In YLD04, we showed that the acceleration by turbulence dominate over other processes and can provide supersonic speed to the grains. On the other hand, it is known (\citealt{Spitzer76, Shull78, Mckee87}, \citealt*{Ellison97}) that grains can also be efficiently accelerated through betatron process by the compressions at shocks. In this paper, we shall consider the betatron acceleration of dust by both incompressible and compressible turbulence. The intent is to provide a comprehensive view of current understanding of grain dynamics in MHD turbulence.

To describe the
turbulence statistics we use the analytical fits to the statistics
of Alfv\'{e}nic modes obtained in \citet*{CLV_incomp} and compressible modes obtained in \citet[hereafter CL02]{CL02_PRL}, which is consistent with the spectra of turbulence velocity from observations \citep{Laz09rev}. In \S2, we provide a brief description of the relevant properties of MHD turbulence. In \S3, we study acceleration by large scale turbulence, where acceleration in both collisionless and collisional regimes are discussed. In \S4, a review of alternative acceleration mechanisms is presented. In \S5, we apply the results to interstellar grains and give a comparison of different mechanisms. Implications of our results are included in the discussion of in \S6, while the summary is provided in \S7.

\section{MHD Turbulence and Its statistical properties}

Turbulent acceleration may be viewed as the acceleration by a spectrum
of MHD waves that can be decomposed into incompressible Alfv\'{e}nic,
and compressible fast and slow modes (see CL02). Separation of MHD perturbations in compressible media into fast, slow and Alfv\'en modes is discussed in GS95, \citet{LG01}, and CL02. The actual decomposition of MHD turbulence into Alfv\'en, slow and fast modes was performed in CL02, who also quantified the intensity of the interaction between different modes (see below).
Unlike hydrodynamic turbulence,
Alfv\'{e}nic turbulence is anisotropic, with eddies elongated along
the magnetic field. This happens because it is easier to mix the magnetic
field lines perpendicular to the direction of the magnetic field rather
than to bend them. As eddies mix the magnetic field lines at the rate
$k_{\bot}v_{k}$ , where $k_\bot$ is the perpendicular component of wavenumber k, $v_{k}$ is the mixing velocity at this scale,
the magnetic perturbations (waves) propagate along the magnetic field
lines at the rate $k_{\parallel}v_{A}$, where $k_\|$ is the parallel component of the wave number k, $v_A$ is the Alfv\'en speed. The Alfv\'{e}nic turbulence
is described by GS95 model. The corner stone of the GS95 model is
a critical balance between these rates, i.e., $k_{\bot}v_{k}\sim k_{\parallel}V_{A}$,
which may be also viewed as coupling of eddies perpendicular to the
magnetic field and wave-like motions parallel to the magnetic field.
From these arguments,
the scale dependent anisotropy $k_{\parallel}\propto k_{\perp}^{2/3}$and
a Kolmogorov-like spectrum for the perpendicular motions $v_{k}\propto k^{-1/3}$
can be obtained. 
It was conjectured in Lithwick \& Goldreich (2001) that the GS95 scaling
should be approximately true for Alfv\'{e}n and slow modes in moderately
compressible plasma. 
For magnetically dominated, the so-called low $\beta$ plasma, CL02
showed that the coupling of Alfv\'{e}n and compressible modes is
weak and that the Alfv\'{e}n and slow modes follow the GS95 spectrum.
This is consistent with the analysis of observational data \citep{Laz09rev} as well as with the
electron density statistics \citep*[see][]{Armstrong95}.
According to CL02, fast modes evolves separately and are isotropic. 

Depending on whether the turbulence velocity on the energy injection scale $L_{inj}$, $\delta V$ is larger or smaller than the Alfv\'en speed, turbulence can be either super-Alfv\'enic ($M_A\equiv (\delta V/v_A)>1$) or sub-Alfv\'enic ($M_A<1$). Turbulence is considered super-Alfv\'enic in intracluster medium. Although less certain, the interstellar turbulence may also be in this regime \citep{Beck01, Padoan04}. 
SuperAlfv\'enic turbulence evolves with hydrodynamic Kolmogorov cascade till the scale $L=L_{inj}(\delta V/v_A)^3\equiv L_{inj}M_A^3$, where the turbulence velocity becomes equal to the Alfv\'en speed $v_A$%
\footnote{Although in general turbulent
generation of magnetic field which will bring kinetic and magnetic
energy to equipartition, the dynamo process takes time and super-Alfv\'enic turbulence can appear as a transient phenomenon.%
}. Below this scale, magnetic field becomes dynamically important and the MHD cascade described above sets in. 

We do not deal with the imbalanced turbulence \citep[see][]{LGS07, BL08, Chandran08} in this paper, as the degree of imbalance in compressible interstellar is unclear. We consider all the turbulent energy is injected from large scales and do not include additional energy injection on small scales arising from compressions of cosmic rays \citep{LazBer06}. Note that the Alfv\'enic turbulence and slow modes are different from waves as they are short lived, only fast modes may resemble wave turbulence.

\section{Acceleration of Grains by Large Scale Turbulence}
\label{main} 

There exists an important
analog between dynamics of charged grains and dynamics of cosmic rays (CRs) and we shall modify the
existing machinery used for cosmic rays to describe charged grain
dynamics%
\footnote{In what follows we assume that the time scale for grain charging is
much shorter than the grain Larmor period. %
}. 

The energy exchange between the particles
and the waves involves both resonant and nonresonant interactions.
Resonant interaction was considered in YL03
 and YLD04. Here we shall be investigating the role of nonresonant interactions on grain dynamics. The role of the non-resonant acceleration of CR by large scale compressions were discussed earlier \citep{Berger, Kulsrud71, Ptuskin88}, and further clarified for MHD turbulence in \citet{CL06}, but there are substantial differences between cosmic rays and grains. In particular, the gyroradii of grain are much larger and can get comparable to the correlation length of turbulence. Moreover, grains are moving much more slowly than the CRs and therefore scattering is not important. Mean free path $l_{mfp}$ of grains are thus much longer and there exists a more important collisionless interaction regime with $l<l_{mfp}$, which we shall demonstrate is the dominant nonresonant interaction in the paper. 

Apart from acceleration,
a grain is subjected to gaseous friction. For the sake of simplicity
we assume here that grains are moving in respect to stationary gas and
turbulence provides nothing but the electromagnetic fluctuations.
The acceleration of grains arising from the gas motion is considered
separately (see LY02). We describe the stochastic acceleration of dust by
the Brownian motion equation \citep*[see][]{Kalmykov96}: $mdu/dt=-u/S+Y$, where $m$ is the
grain mass, $u$ is the grain speed, $Y$ is the stochastic acceleration force, $S=t_{drag}/m$
is the mobility coefficient \citep{Draine85, LY02}.  The drag time due to collisions with atoms is essentially
the time for collisions with the amount of gas with the mass equal
to that of the grain, $t_{drag}^{0}=(a\rho_{gr}/n_{n})\sqrt{\pi/(8m_{n}k_{B}T)}$,
where $a$ is the grain size, $T$ is the temperature, $n_n$ is the neutral density, $\rho_{gr}$
is the mass density of grain. We adopt $\rho_{gr}=3.8$gcm$^{-3}$
for silicate grains. The ion-grain cross-section due to long-range
Coulomb force is larger than the atom-grain cross-section. Therefore,
in the presence of collisions with ions, the effective drag time decreases
by the factor \citep{DraineSal79},
\begin{eqnarray}
\alpha&=&1+\frac{n_{\rm H}}{2n_n}\sum_{i}x_i\left(\frac{m_{i}}{m_n}\right)^{1/2}\sum_{Z}f_Z\left(\frac{Ze^{2}}{ak_BT}\right)^{2} \nonumber \\
& &\ln\left[\frac{3(k_B
T)^{3/2}}{2e^{3}|Z|(\pi xn_{\rm H})^{1/2}}\right].
\end{eqnarray}
where $n_H$ is the hydrogen density, $x_{i}$ is the abundance, relative to hydrogen, of ion $i$ with mass $m_{i}$, $x=\sum_i x_i$, and $f_{Z}$ is the probability of the grain being in charge state $Z$. When the grain velocity
gets supersonic \citep{Purcell69}, the gaseous drag time is given by $t_{drag}^{s}=t_{drag}/(0.75u/c_S+0.75c_{S}/u-c_{S}^{2}/2u^{2}+c_{S}^{3}/5u^{3})$,
where $c_{S}$ is the sound speed.

Multiply the above Brownian equation above by $u$ and take the ensemble
average, we obtain

\begin{equation}
m\frac{d<u^{2}>}{dt}=-\frac{<u^{2}>}{S}+<\dot{\epsilon}>,\label{diff}\end{equation}
 Following the approach similar to that in \citet{Melrose80}, we can
get from Eq.(\ref{eq:Fokker}) the energy gain rate $<\dot{\epsilon}>$
for the grain

\bea
<\dot{\epsilon}>&=&\frac{1}{4p^{2}}\frac{\partial}{\partial p}(up^{2}D_{p})\sim\frac{D_p}{m},
\label{epsilon}
\eea
where $D_p$ is the momentum diffusion coefficient, and shall be derived below for different circumstances.

\subsection{Acceleration in incompressible turbulence}

One important nonresonant process is betatron acceleration. It arises from the compression of the magnetized fluid, which induces nonzero curl of electric field $\nabla\times {\bf E}$ in the plane perpendicular to the local magnetic field. The same betatron process was studied earlier in the context of shock acceleration, where grains are accelerated at postshock due to the compression. Here we shall investigate the betatron acceleration in turbulence. We stress that the betatron process operates even in incompressible turbulence, where magnetic field can be compressed by pseudo-Alfv\'en modes, which are the incompressible limit of slow modes.


Let us start with incompressible turbulence. The average electric field during one gyro-orbit can be estimated as \citep{CL06},

\bea
{\bar E}&=&\frac{1}{2\pi r_g}\int (\nabla \times {\bf E})\cdot d{\bf s}, \nonumber \\
&=&\frac{1}{2\pi r_g c}\int \left[\nabla \times ({\bf v}\times {\bf B})\right]\cdot d{\bf s}, \nonumber \\
&\approx&\frac{1}{2\pi r_g c}\int({\bf B\nabla}\cdot {\bf v}-{\bf B_0}\cdot {\bf \nabla v})\cdot d{\bf s},
\label{betaelectric}
\eea
where $r_g$ is the Larmor radius of grain, c is the light speed. The first term arises from the compression of the fluid, which is equal to zero in incompressible fluid. We approximate the total magnetic field ${\bf B}$ by the mean field ${\bf B}_0$ in the second term. We also used the fact that the MHD wave in incompressible medium is transverse, i.e., ${\bf v}\perp{\bf k}$. Therefore we focus here on the second term. In this case,

\be
{\bar E}\approx -B_0k_\|v_{k,\|}r_g/2c.
\ee
, where $v_{k,\|}$ is the fluid velocity parallel to the magnetic field. The rate of momentum change is then

\be
\frac{dp_\bot}{dt}=q{\bar E}\sim qB_0k_\|v_{k,\|}r_g/2c =p_\bot k_\|v_{k,\|}/2. 
\label{acceleration}
\ee
where $q=Ze$ is the charge of grain and $p_\bot$ is the component of grain momentum perpendicular to the magnetic field. The acceleration is canceled out by the adiabatic loss in the parallel direction if the particle velocities are isotropic. Thus it only operates on the scale less than the mean free path $l_{mfp}$ \citep{CL06}, from which we get 

\be
D_p\sim \left(\frac{dp_\bot}{dt}\right)^2{\Delta t} \sim p_\bot^2 (k_\|v_{k,\|})^2/4\times \Delta t \sim\frac{p^2_\bot V^2}{6L{\rm max}(v_A, u)}
\label{eq:Fokker}
\ee
where we adopted the GS95 scaling for the pseudo-Alfv\'en modes, $v_k\sim V (k_\bot L)^{-1/3}$ and $k_\|\sim k_\bot^{2/3}L^{-1/3}$,
 $V$ is the turbulence velocity at the injection scale L of MHD turbulence. For pseudo Alfv\'en modes, ${\bf v}_k\perp {\bf k}$ and resides in the ${\bf k-B}$ plane. Morever most energy is concentrated in the modes with ${\bf k}\perp {\bf B}$ in the GS95 model. Therefore we used $v_{k,\|}\simeq v_k$ in the above equation. Unlike cosmic rays, $\Delta t =1/[k_\|{\rm max}(v_A,u)]$ is determined by the minimum of the eddy turnover time and the particle streaming time. This is because the scattering of grains are infrequent because of the low speed as pointed out in YL03.

\begin{figure*}
\includegraphics[width=0.95\textwidth,
  height=0.28\textheight]{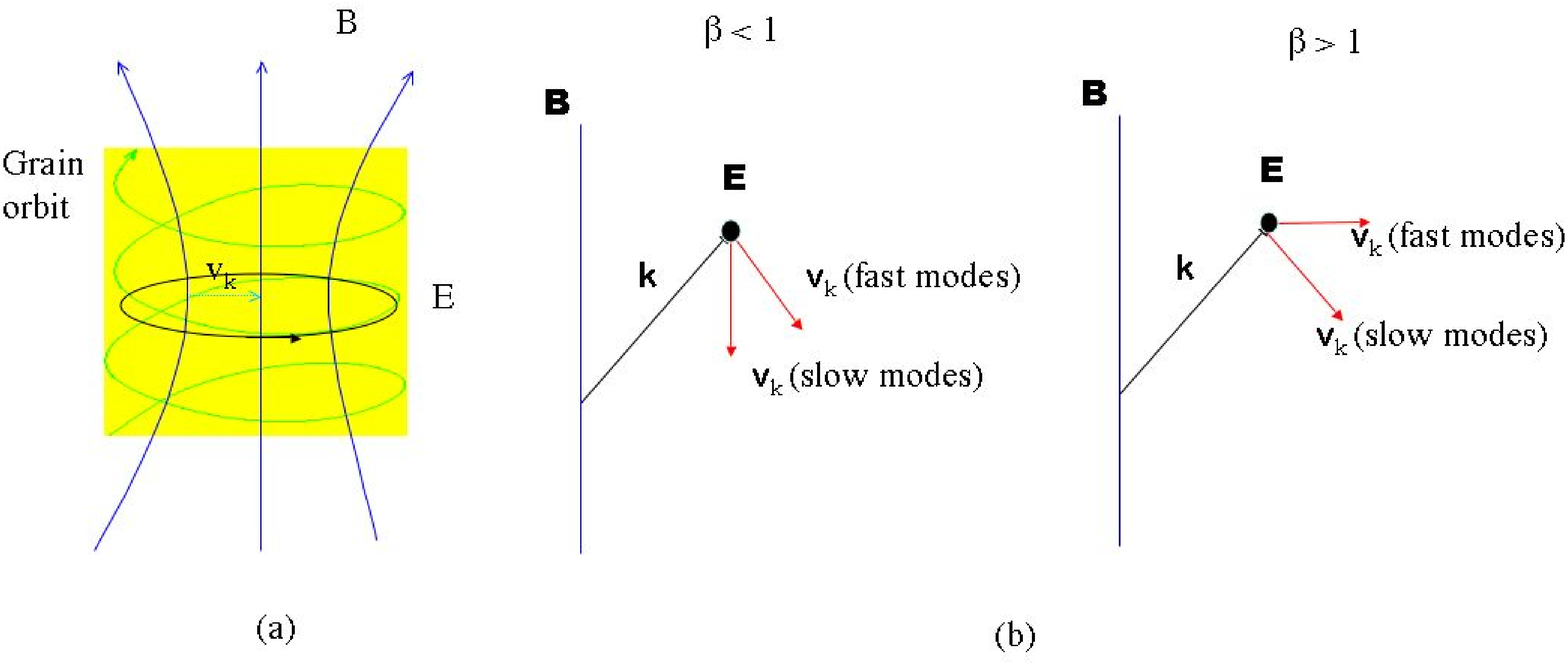}
\caption{(a): the compression of magnetic field induce a nonzero curl of electric field, which accelerate grains as grains are orbiting around the magnetic field; (b): the direction of velocity perturbation and electric field induced by compressible modes.}
\end{figure*}
 
\subsection{Acceleration in compressible turbulence}
 
In compressible medium, there are both slow modes and fast modes. For compressible modes, the above mechanism also operates on the scale $l<l_{mfp}$. The compression on large scale $l>l_{mfp}$ is much less efficient in accelerating grains for the following reason. Since usually $l_{mfp}\gg r_g$, $l>l_{mfp}$ indicates $l\gg r_g$. Lack of sufficient scattering the adiabatic invariant of grain $v_\bot^2/B$ is conserved in this case and the change of grain velocity is only a factor of few \citep{CL06}. 

The betatron acceleration by fluctuations on $l<l_{mfp}$ does not work for small grain, whose mean free path is less than the turbulence damping scale $l_c$.  Lack of scattering, the mean free path of dust is usually determined by the gaseous or plasma drag $l_{mfp}=u t_{drag}$.

Let us consider the betatron acceleration by the compressible modes on the small scales $l<l_{mfp}$. The first term in Eq.(\ref{betaelectric}) is nonzero and needs to be accounted for. 

\bea
{\bar E}&=&\frac{1}{2\pi r_g}\int (\nabla \times {\bf E})\cdot d{\bf s}, \nonumber \\
&=&\frac{1}{2\pi r_g c}\int \left[\nabla \times ({\bf v}\times {\bf B})\right]\cdot d{\bf s}, \nonumber \\
&\approx&\frac{k_\bot v_{k,\bot} B_0r_g}{2c},
\eea
where $v_{k,\bot}$ is the fluid velocity perpendicular to the magnetic field. Accordingly, the electrical force and the momentum diffusion coefficient are 

\be
\frac{dp}{dt}=q{\bar E}=k_\bot  v_{k,\bot} p_\bot/2,
\ee
\be
D_p=\left(\frac{dp}{dt}\right)^2\Delta t\sim k_\bot^2p_\bot^2 v^2_{k,\bot}{\rm min}(t_{wav},t_{gc})/4,
\ee
where $t_{wav}$ is the wave period and $t_{gc}$ is the confinement time of the grains.

For {\it slow} modes, 
\be
D_{p,k}=\frac{k_\bot^{2/3}L^{-1/3}p_\bot^2V^2}{4{\rm max}(v_{ph},u)}<\frac{v^2_{k,\bot}}{v^2_k}>,
\label{beta_slow}
\ee
where $v_{ph}={\rm min}(v_A,c_S)$ is the phase speed of the slow modes. The velocity displacement of slow modes is in the direction of $\hat{\bf k}_\|-\frac{\beta}{2}\sin\theta\cos\theta~ \hat{\bf k}_\bot$ and $\bot~ {\bf k}$ in the high $\beta$ medium \citep{CL06}, where $\theta$ is the wave pitch angle, i.e., the angle between the wave vector {\bf k} and {\bf B}, $\beta=2c_S^2/v_A^2$ is the ratio of thermal pressure to magnetic pressure. Therefore, $<v^2_{k,\bot}/v^2_k>\approx <\cos^2\theta>$ in high $\beta$ medium and $<v^2_{k,\bot}/v^2_k>\approx \beta^2<\sin^2\theta\cos^2\theta>/4$ in low $\beta$ medium, Because of the strong anisotropy ($k_\bot\gg k_\|$), $<\sin^2\theta>\approx 1$, and $<\cos^2\theta>\approx (k_\|/k_\bot)^2=(k_\bot L)^{-2/3}$. All the scales from $l_{mfp}$ to the truncation scale $l_c$ contribute to the acceleration. Since the contribution from the scales smaller than the gyroradius has been included in the treatment of gyroresonance in YL03, we only account for the eddies larger than gyroradii. Integrating Eq.(\ref{beta_slow}), we get

\be
D_p\sim \frac{p^2V^2}{6L{\rm max}(v_{ph},u)}{\rm min}\left(\frac{\beta^2}{4},1\right)\ln\left[\frac{l_{mfp}}{{\rm max}(l_c, r_g)}\right] 
\ee   

In low $\beta$ medium, the slow modes cause marginal compression of magnetic field with the velocity perturbations being nearly parallel to the magnetic field. That is why the acceleration by the slow modes in the low $\beta$ regime is substantially reduced.

For {\it fast} modes, the velocity perturbation is $\bot~{\bf B}$ in the low $\beta$ case and $\|~ {\bf k}$ in the high $\beta$ case and therefore $v^2_{k,\bot}/v^2_k\approx 1$ for $\beta<1$ and 2/3 for $\beta>1$ since fast modes are isotropic. Unlike slow modes, the cascade of fast modes is slower, namely, the cascading time is longer than the wave period. Therefore, one needs to take into account periodic motion of the fast modes. In case when particles are confined in the eddy longer than the wave period, the expansion ensuing the compression will take the energy back from the particles. Therefore, the acceleration only happens if the particles leave the eddies within one compressional period. Since grains experience marginal scattering,    
it indicates that only super-Alfv\'enic grains ($\beta<1$) or supersonic grains ($\beta>1$) can be accelerated. 
\be
D_p=\left(\frac{dp}{dt}\right)^2\Delta t\sim k_\bot^2p_\bot^2 v^2_{k,\bot}t_{gc}/4\sim k^{1-2\xi} L^{-2\xi} p_\bot^2V^2/(6u),
\ee
where we have used $v_k=(kL)^{-\xi}$, $\xi=1/3$ for Kolmogorov turbulence and $1/4$ for acoustic turbulence. Unless otherwise specified, we adopt $\xi=1/4$ to enable comparison with our  earlier results (YL03, YLD04).
Since it increases with the wave number k, the contribution on the smallest possible scale is dominant.  Excluding the contributions from scales $\la r_g$, the corresponding scale is then max$(r_g,l_c)$. Accordingly,

\be
D_p\sim \frac{p^2V^2}{9u{\rm max}(r_g,l_c)^{1-2\xi}L^{2\xi}}.
\label{beta_fast}
\ee

Comparing the above equations, we see that usually the acceleration by fast modes is much more efficient than by slow modes. Low speed grains, however, can only be accelerated by the slow modes for the reason we discussed above.

\section{Alternative Acceleration mechanisms}
\label{review}
\subsection{Gyroresonance}

Gyroresonance acceleration of charged grains by a spectrum
of MHD waves decomposed into incompressible Alfv\'{e}nic,
and compressible fast and slow modes (see CL02) was first described
in \citet{YL03}. Gyroresonance occurs when
the Doppler shifted frequency of the wave in the grain's guiding center
rest frame $\omega_{gc}=\omega-k_{\parallel}u_\|=n\Omega$ is a multiple of
the grain gyrofrequency, $u_\|$ is the component of grain velocity parallel to the local magnetic field.
For low speed grains, we only need to consider the resonance at $n=1$. From the resonance condition, we see that gyroresonance requires perturbations on the Larmor scale and therefore only applicable to large grains whose gyroradii are larger than the dissipation scale of the turbulence $l_c$. The gyroresonance
changes both the direction and absolute value of the grain's momentum
(i.e., scatters and accelerates the grain). The ratio of the scattering and acceleration rates is approximately $u^2/v^2_A$. For low speed grains, therefore, acceleration is the main effect.

\subsection{Transit time damping}

In the calculations in \S\ref{main}, we neglected the mirror force $-\mu\nabla B=-mu^2_\bot\nabla B/(2B)$, which scatters the grains, but does not increase the energy. In the small amplitude limit $\delta B\ll B$, particles should keep in phase with the wave $k_\|v_\|=\omega$ in order to have effective interactions with the compressions of magnetic field. This process is often referred as transit time damping (TTD) in the literature \citep[see, e.g.,][]{Achterberg, SchlickeiserMiller, YL04}. In this case, the mirror force redirect the energy gain in the perpendicular direction to the parallel direction without changing the total grain speed. In fact, our test calculation of the TTD using quasilinear approach is consistent with the results in \S\ref{main} within order of unity. I therefore do not think that in terms of
the acceleration TTD is different from the betatron process discussed in this paper.

\subsection{Hydrodynamic drag}
In hydrodynamic turbulence, the grains motions are caused by the frictional interactions with the gas. On large scales grains are coupled with the ambient gas,
and the slowing fluctuating gas motions will only cause an overall advection of the grains with the gas (Draine 1985), which we are not
interested. On small scales
grains are decoupled. The largest velocity difference occurs on the largest
scale where grains are still decoupled. Thus the characteristic
velocity of a grain with respect to the gas corresponds to the velocity
dispersion of the turbulence on the time scale $t_{drag}$. In the MHD case, the grain perpendicular motions
are constrained by the Larmor gyration unless $t_{drag}\Omega<1$ \citep{LY02}. 

The grain motions get modified when the damping time scale of the turbulence $\tau_{c}$ is longer than either $t_{drag}$
or $1/\Omega$. In this case, a grain samples only
a part of the eddy before gaining the velocity of the ambient gas. The largest shear appears on the damping scale, and thereby $u\sim v_c {\rm min}(1/\Omega,t_{drag})/\tau_c$, where $v_c$ is the velocity on damping time scale $\tau_c$.

\subsection{Ponderomotive force}

A cross-field acceleration by the ponderomotive force was suggested by \citet{Shukla} as an another independent mechanism in turbulence to provide grains with supersonic speeds. If this were true, it would indeed indicate a dramatic shift in the whole scenario, especially for small grains. We find however their formalism is erroneous. Let us start from the electric field that could arise in the varying magnetic field. Considering the momentum equation of the electron, where the electron inertia can be neglected in the low frequency MHD regime,

\be
n_ee{\bf E}=-\frac{n_ee}{c}{\bf v}_e\times {\bf B}-\gamma_eT_e\nabla n_e,
\ee
where $n_e,\,{\bf v}_e,\,T_e,\,\gamma_e$ are the density, velocity, temperature and adiabatic index of electrons. Combining it with the quasi-neutrality condition $n_e=n_i+Zn_d$ and the induction equation,
\be
\nabla \times {\bf B}=\frac{4\pi e}{c}(n_i{\bf v}_i+Zn_d{\bf u}-n_e{\bf v}_e),
\ee
we obtain

\be
{\bf E}=\frac{1}{n_ee}\left(\frac{\nabla\times {\bf B}}{4\pi}-n_i{\bf v}_i-Zn_du\right)\times {\bf B}-\gamma_eT_e\nabla n_e
\ee
where $n_i,\,{\bf v}_i$ are the density and velocity of ions. If we neglect the thermal term in the low $\beta$ medium we consider, the electric field is perpendicular to the magnetic field and cannot accelerate the particles longer than the Larmor period. The net effect of the electric field is to induce a drift in the direction ${\bf E}\times {\bf B}$. The corresponding drift speed is 

\be
{\bf u}=c\frac{{\bf E}\times {\bf B}}{B^2}={\bf v}_i-\frac{c{\bf k}\times \delta {\bf B}}{4\pi n_ee}
\ee

Since the acceleration is on the time scale of Larmor period, the wavenumber in the above equation is given by $k\approx r_g^{-1}$ and ion velocity that should be adopted is the velocity of the ions (gas) on the Larmor scale. It turns out then the second term is much less than the first term in the above equation, and the drift is in fact the same process as the gaseous drag we already accounted for in \citet{LY02} and also the Fig.1 in this paper. In other words, there is no such an additional ponderomotive acceleration as discussed in \citet{Shukla}.

\subsection{Weak shocks}
Simulations indicate that the density structures in supersonic MHD turbulence are associated with slow shocks \citep*{BLC05}. Therefore in supersonic turbulence, one also needs to take into account the contribution of grain acceleration from shocks \citep{Epstein80, Mckee87, Ellison97}. The shocks in the supersonic turbulence require better quantitative description, nevertheless.

\subsection{$H_2$ rocket and radiative pressure}

A different mechanism of driving grain motions is a residual imbalance
in {}``rocket thrust{}'' between the opposite surfaces of a rotating
grain \citep{Purcell79}. This mechanism can provide grain relative motions
and preferentially move grains into molecular clouds. Three causes for the thrust were
suggested by \citet{Purcell79}: spatial variation of the accommodation
coefficient for impinging atoms, photoelectric emission, and H\( _{2} \)
formation. The latter was shown to be the strongest among the three.
The uncompensated force in this case arises from the difference of
the number of catalytic active sites for H\( _{2} \) formation on
the opposite grain surfaces. The nascent H\( _{2} \) molecules leave
the active sites with kinetic energy \( E \) and the grain experiences
a push in the opposite directions. $H_2$ thrust potentially may be important for small grains. In the typical interstellar environment, small grains flip frequently according to recent study (Hoang \& Lazarian 2009). As a result, the thrusts are smeared out.

Dust grains exposed to anisotropic interstellar radiation fields are subjected to forces as well as torques \citep{Purcell79, WD01a}.
These forces arise from photoelectric emission, photodesorption as well as radiation
pressure. \citet{WD01a} considered the forces and calculated the drift velocity for grains of different
sizes. The velocities resulted from these processes, however, are smaller than those arising from interactions with turbulence in the interstellar medium (YLD04). 

\section{Results}
\label{result}

\begin{figure*}
\begin{center}
\includegraphics[width=0.45\textwidth,
  height=0.25\textheight]{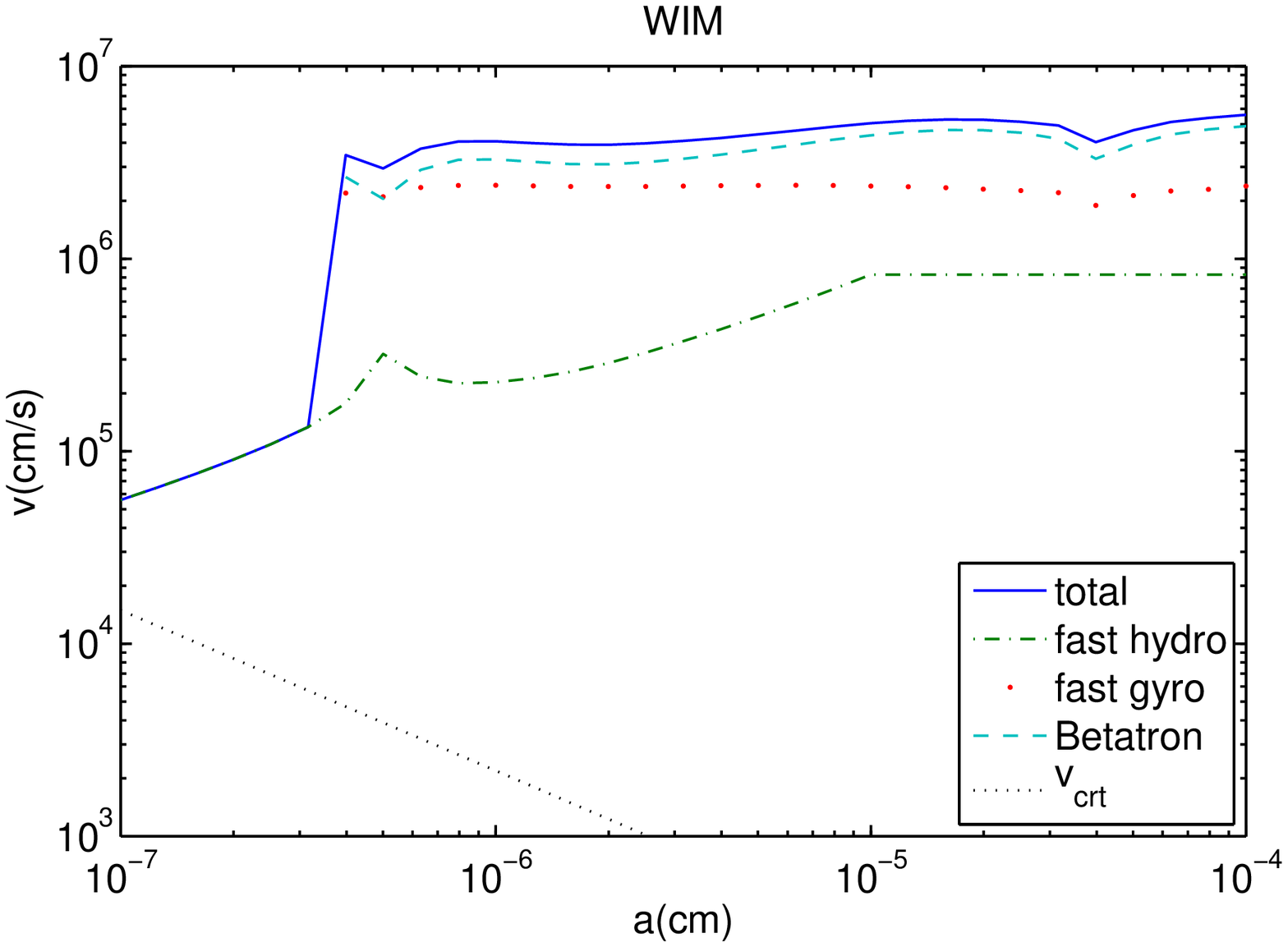}
\includegraphics[width=0.45\textwidth,
  height=0.25\textheight]{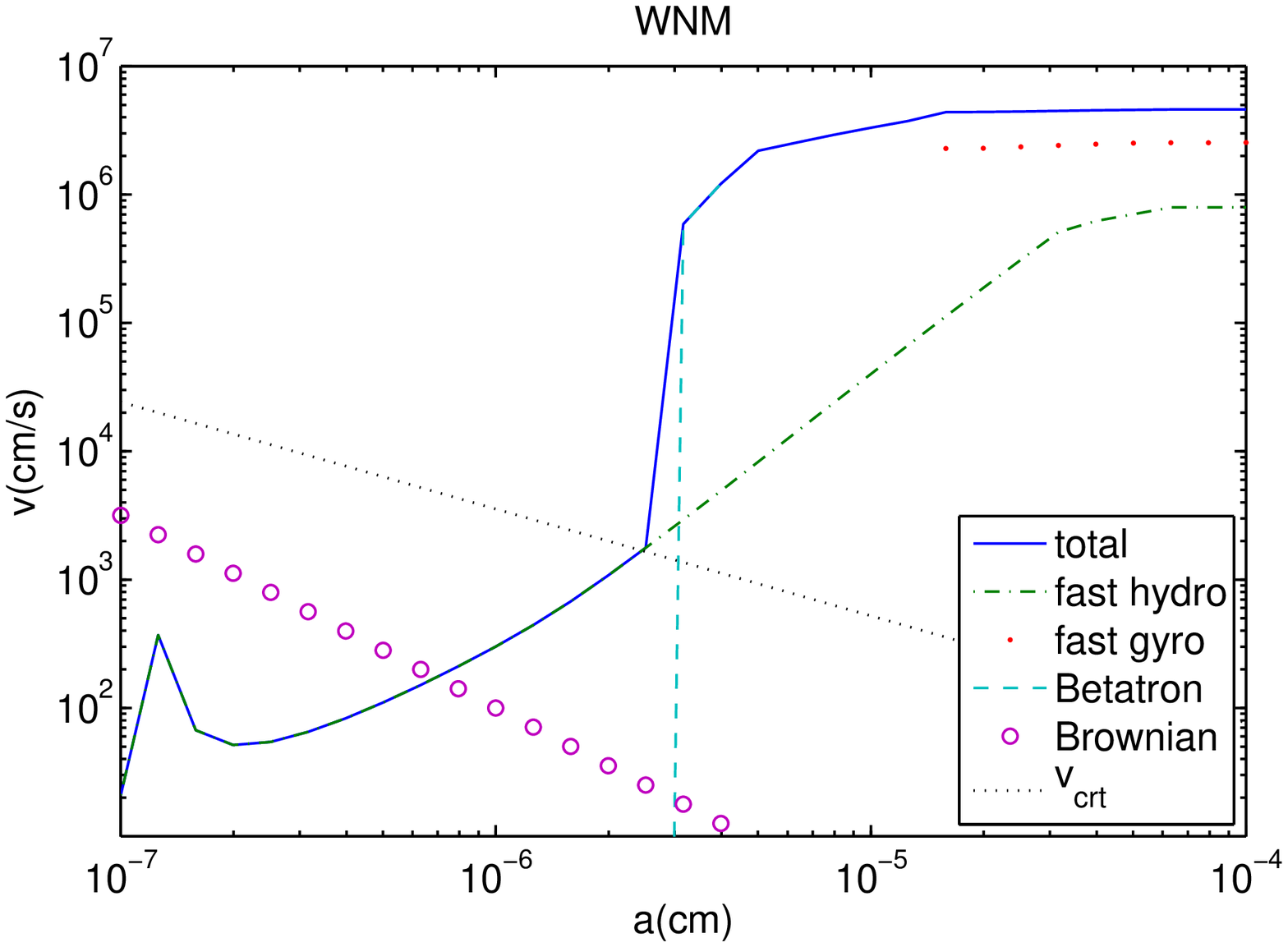}
\end{center}
\caption{Grain velocity vs. size owing to different acceleration processes in WIM ({\it left}) and WNM ({\it right}).
 The result from 
betatron acceleration is represented by the dashed line. The '.' line represents the gyroresonance with fast modes. The
dash-dot line is the result from hydro drag with fast modes. The dotted line refers to the threshold velocity for coagulation. Contributions from 
different processes are approximately additive as shown by the solid line.}
\label{WIM}
\end{figure*}

To compare with earlier work (YLD04), we consider first the warm ionized medium (hereafter WIM) with the same parameters, $T=8000$K, $n_{e}=0.1{\rm cm}^{-3}$,
$B_{0}=3.35\mu$G (see table\ref{para}). Here, we consider grains in the range of $10^{-7}$cm$<a<10^{-4}$cm,
which carry most grains mass in ISM. We adopt the mean grain
charge as given in YLD04, which was obtained by balancing the collisions with electrons
and collisions with ions as well as photoelectric emission \citep{WD01b}. 
We assume
that for MHD turbulence the injection of energy happens at the scale $L$ where the equipartition
between magnetic and kinetic energies, i.e., $V=V_{A}$, is reached.
If we adopt a velocity dispersion $\delta V=40$km/s at the scale $l=30$pc, then turbulence in WIM is super-Alfv\'{e}nic at
this scale. If turbulence at this regime follows hydrodynamic cascade,
then $L$ is 3.75pc where the injection velocity is $V=V_{A}=20$km/s. The acceleration is dominated by the interactions with fast modes. Since fast modes are long lived and can only accelerate particles whose crossing time is shorter than the wave period, the betatron acceleration only operates with fast-moving large grains. The result for the WIM is demonstrated in Fig.\ref{WIM}, where the outcome from different grain acceleration processes are compared. 

Next we consider a warm neutral medium (hereafter WNM), with T=6000K, $n_H=0.3 {\rm cm}^{-3}$, $B_0=5.8\mu G$ and the same turbulence injection velocity and scale as in WIM (see table 1 for other parameters). In partially ionized medium, a viscosity
caused by neutrals results in decoupling on the characteristic time
scale \citep{LY02} $
t_{damp}\sim\nu_{n}^{-1}k^{-2}\sim(l_{n}v_{n})^{-1}k^{-2},\label{cutoff}$
 where $\nu_{n}$ is the kinetic viscosity, $l_{n}$ is the neutral
mean free path, $v_{n}$ is the thermal velocity of neutrals. When its cascading
rate $\tau_{k}^{-1}$ equals to the damping rate %
 $t_{damp}^{-1}$, turbulence is considered
damped. This defines the truncation scale of the turbulence $l_c$. $\tau_k=kV^2/v_{ph}$ for fast modes \citep{CLV_incomp} and $\tau_k=k_\bot^{2/3}L^{-1/3}v_A$ for Alfv\'en and slow modes \citep{GS95}. Then for WNM, one gets $l_{c}\simeq8.3\times10^{15}$cm for the fast modes and $l_c\simeq5.3\times 10^{17}$cm for Alfv\'en and slow modes. This introduces a cutoff grain size for the acceleration by fast modes, for which the Larmor period is equal to the damping time scale.   The result for grain acceleration is shown in Fig.\ref{WIM}. As we see, the betatron acceleration operates for an extended range of grains  and can be more efficient compared to gyroresonance. In particular, fast modes accelerate further the super-Alfv\'enic grains and slow modes dominant the acceleration for the smaller grains.

In YL03, we proposed gyroresonance as one important acceleration mechanism for grain dynamics. We showed that the magnetostatic approximation for turbulence is invalid for grains, whose velocities are less or comparable to the velocity perturbations in turbulence. As a result, the resonance function in the quasi-linear theory is broadened from the resonance wavenumber $k_{\|,res}\sim 1/r_g$ to all k in the inertial range. To cut off the spurious resonance contribution from the large scale perturbations, we truncated the integration for the momentum diffusion at the $k_{\|,res}$ and discarded all the interactions from the larger scales. A complementary study is conducted in this paper to include the acceleration from the large scales. Our results show that the nonresonant interaction is comparable to the gyroresonance and can get even more efficient in super-Alfv\'enic  turbulence and the strength of magnetic field.       

 

\begin{table*}
\label{para}
\begin{tabular}{ccccccccc}
\hline
\hline
&$T(K)$&$n_H({\rm cm}^{-3})$&$n_e({\rm cm}^{-3}$)&$G_{UV}$&B($\mu G$)&L(pc)&$V=v_A({\rm km \cdots}^{-1}$)&$l_c({\rm cm}$)\\
\hline
WIM&8000&0.1&0.0991&1&3.35&3.7&20&...\\
WNM&6000&0.3&0.03&1&5.8&3.7&20&$8.3\times 10^{15}$\\
\hline
\hline
\end{tabular}
\caption{The parameters of idealized ISM phases and relevant damping. 
Among them, $n_H$ is the number density of H, $n_e$ is the number density of 
electron, $G_{UV}$ is the UV intensity scale factor, V is the injection velocity.
L is the injection scale of fast modes, which is taken as the scale where the turbulence velocity is equal to the Alfv\'en speed (see the text for details). The
dominant damping mechanisms for fast modes are given with the corresponding damping timescale $\tau_c$. WNM=warm neutral
medium, WIM=warm ionized medium. }
\end{table*}

Our results show that only large grains can be effectively accelerated by the large scale compressions in turbulence. It is due to the fact that the scattering of grains is marginal. With the ballistic motions, the adiabatic invariant is conserved provided that the grain gyroradii are much smaller than the turbulence injection scale. The grain energy gain is thus limited. For large grains, their gyroradii are close to perturbation scale and there is no such a constraint. Our results are therefore self-consistent.   


How would the results vary as the parameters of the partially ionized
medium vary? The relative importance of the
gyroresonance and betatron acceleration depends on the magnitude of the magnetic
field as well as the intensity of turbulence. With the increase of magnetic field, the relative efficiency of betatron acceleration decreases compared to gyroresonance. On the other hand, the betatron acceleration increases faster with the intensity of turbulence than gyroresonance. More specifically, gyroresonance dominates in equipartition turbulence and sub-Alfv\'enic turbulence ($M_A=\delta V/v_A\lesssim 1$).  In super-Alfv\'enic turbulence ($M_A=\delta V/v_A>1$), betatron acceleration becomes important and works for extended range of grains being less constrained by the turbulence damping. In \S\ref{result}, for instance, we adopt $\delta V=40$km/s at $l=40$pc. If one takes $\delta V=20$km/s as in YLD04, the nonresonant and resonant acceleration rates are comparable. It is therefore sufficient to consider only the gyroresonance. Although turbulent
generation of magnetic field tend to bring kinetic and magnetic
energy to equipartition, the time required is long (a few crossing time at the injection scale according to \citealt{CVBL08}), during which grains can be accelerated through the betatron process with the super-Alfv\'enic turbulence. Even if the velocity dispersions that we adopt in table.1 are excessive for the entire ISM, the local driving may bring
them to such and even larger values.

 
\section{Discussions}

This paper intends to encompass all possible situations where dust grains can be accelerated. As presented in \S\ref{review}, various processes have been proposed in the literature for the grain acceleration. In regard to the plasma processes alone, there have been shock acceleration \citep[e.g.,][]{Epstein80, Mckee87}, gyroresonance\citep{YL03}, ponderomotive force \citep{Shukla}. We intend to supplement the list by taking into account nonresonant betatron acceleration in MHD turbulence. This process was discussed in the context of cosmic ray acceleration \citep[e.g.,][]{Ptuskin88, CL06}, but has not been considered for charged grains. The study here demonstrates that the betatron acceleration can be important in super-Alfv\'enic turbulence.    

The steady state for the strongly super-Alfv\'enic turbulence is not achievable due to dynamo generation. However, numerical simulations \citep[see][]{CVBL08} show that the steady state may not necessarily have the exact equipartition and kinetic energy may sometimes dominate. In fact, even in the situation when turbulence in terms of its mean velocity dispersion in the volume is sub-Alfv\'enic, density fluctuations may induce super-Alfv\'enic turbulence. This is shown in \citet{BurkKowal08}.

It has been shown that the composition of the galactic cosmic ray
seems to be better correlated with volatility of elements \citep*{Ellison97}. The more refractory elements are systematically
overabundant relative to the more volatile ones. This suggests that
the material locked in grains must be accelerated more efficiently
than gas-phase ions \citep*{Epstein80, Ellison97}.
The stochastic acceleration of grains, in this case, can act as a
preacceleration mechanism. The accelerated grains are more efficiently scattered by MHD perturbations, which is a key component for diffuse shock acceleration.

Grains moving supersonically can efficiently vacuum-clean heavy
elements as suggested by observations \citep{WakkerMathis00}. Grains
can also be aligned if the grains get supersonic (see review by \citealt{Lazarian07rev}). The supersonic grain motions will result in grain alignment with long
axes perpendicular to the magnetic field. In fact, for irregular grains with nonzero degree of helicity, even subsonic motions can get grains aligned as recently illustrated by \citet{LH07}. The critical velocity 
is, however, still unclear. If taking into account mirror reflection, the energy gain of grains is not strictly in the perpendicular direction. Nevertheless, fast moving grains are aligned with the long axes perpendicular to the magnetic field regardless of its direction of motions according to recent studies \citep{LH07}.


In our recent paper \citet{HY09}, we calculated the equilibrium grain size distribution based on our earlier result (YLD04) and obtained corresponding extinction curve in various phases of ISM. There we applied the gyroresonance, which is dominant in equipartition turbulence and subAlfv\'enic turbulence. We believe the bulk of interstellar medium has $M_A\lesssim 1$ because of the dynamo process, the basic conclusion there is valid therefore. In particular environments where turbulence gets super-Alfv\'enic, the process discussed in this paper should be accounted for. As I discussed earlier, the betatron acceleration is most efficient for large grains ($>$ a few tenths $\micron$), so the effect is mainly on shattering. Depending on the intensity of turbulence and the strength of magnetic field, the velocity can be increased by a factor of a few through the betatron acceleration, which could decrease the upper limit of the grain size in the warm media by a factor of few.
In cold media, including cold neutral medium and the molecular clouds, the betatron acceleration is less efficient, for two reasons. First, the low levels of UV and low temperatures result in
reduced grain charge.
Secondly, because of the increased density, the frictional drag is increased. The betatron process is subdominant unless in highly super-Alfv\'enic turbulence, which could appear in a transient state.
 

\section{SUMMARY}

We calculate the acceleration of charged grains arising from betatron acceleration by MHD turbulence. We used the CL02's results for the description
of MHD turbulence, according to which Alfv\'{e}nic turbulence and slow modes follows
GS95 scaldings while fast modes are isotropic. We obtained estimates of grain
velocities in interstellar medium. We showed that
\begin{enumerate}
\item MHD turbulence dominate grain acceleration in interstellar medium.

\item The relative importance of betatron acceleration and gyroresonance depends on the intensity of turbulence. While gyroresonance dominates in sub-Alfv\'enic turbulence, betatron acceleration can be important in super-Alfv\'enic turbulence.

\item Nonresonant acceleration with fast modes are important for large grains in warm media. Slow modes can accelerate sub-micron grains through the nonresonant interactions. 

\item Grain velocities depend on the local conditions of turbulence.

\item Dust gets supersonic via interactions with interstellar turbulence.
The velocities obtained are sufficiently high to
be important for shattering large grains and efficiently absorbing
heavy elements from gas. The acceleration can also
result in mechanical alignment of grains perpendicular to the magnetic field.

\end{enumerate}

\section*{Acknowledgments}
I thank A. Lazarian for reading the manuscript and valuable comments and suggestions. Helpful discussions with J. Cho is acknowledged. This work is supported by Arizona Prize Fellowship.

\bibliography{yan}
\bibliographystyle{apj}

\label{lastpage}

\end{document}